\documentclass[]{aastex631}

\graphicspath{{./}{figures/}}

\begin{document}

\title{Solar Energetic Particle Events and Associated Type II Radio Bursts from Different Source Regions}

\author[0009-0009-7015-0024]{Xuchun Duan}
\affiliation{State Key Laboratory of Solar Activity and Space Weather,National Astronomical Observatories, Chinese Academy of Science, Beijing
100101, People's Republic of China}
\affiliation{School of Astronomy and Space Science,
University of Chinese Academy of Sciences,
Beijing 100049, People's Republic of China}
\author[0000-0001-6655-1743]{Ting Li}
\affiliation{State Key Laboratory of Solar Activity and Space Weather, National Space Science Center, Chinese Academy of Sciences, Beijing 100190, People’s Republic of
China
}
\affiliation{State Key Laboratory of Solar Activity and Space Weather,National Astronomical Observatories, Chinese Academy of Science, Beijing
100101, People's Republic of China}
\affiliation{School of Astronomy and Space Science,
University of Chinese Academy of Sciences,
Beijing 100049, People's Republic of China}

\author[0009-0009-9306-4022]{Yingli Cui}
\affiliation{School of Space Science and Physics, Institute of Space Sciences, Shandong University, Weihai, Shandong 264209, People’s Republic of China}

\author[0000-0002-9534-1638]{Yijun Hou}
\affiliation{State Key Laboratory of Solar Activity and Space Weather,National Astronomical Observatories, Chinese Academy of Science, Beijing
100101, People's Republic of China}
\affiliation{School of Astronomy and Space Science,
University of Chinese Academy of Sciences,
Beijing 100049, People's Republic of China}
\author[0000-0001-7693-4908]{Chuan Li}
\affiliation{School of Astronomy and Space Science, Nanjing University, Nanjing 210023, People's Republic of China}
\author[0000-0001-6344-6956]{Nicolas Wijsen}
\affiliation{Center for mathematical Plasma Astrophysics, KU Leuven, Kortrijk/Leuven, Belgium}

\author{Zelong Jiang}
\affiliation{School of Space Science and Physics, Institute of Space Sciences, Shandong University, Weihai, Shandong 264209, People’s Republic of China}

\author{Yihua Yan}
\affiliation{State Key Laboratory of Solar Activity and Space Weather, National Space Science Center, Chinese Academy of Sciences, Beijing 100190, People’s Republic of
China
}
\affiliation{State Key Laboratory of Solar Activity and Space Weather,National Astronomical Observatories, Chinese Academy of Science, Beijing
100101, People's Republic of China}
\affiliation{School of Astronomy and Space Science,
University of Chinese Academy of Sciences,
Beijing 100049, People's Republic of China}
\author[0000-0002-5431-6065]{Suli Ma}
\affiliation{State Key Laboratory of Solar Activity and Space Weather, National Space Science Center, Chinese Academy of Sciences, Beijing 100190, People’s Republic of
China
}
\author[0000-0001-5657-7587]{Zheng Sun}
\affiliation{School of Earth and Space Sciences, Peking University, Beijing 100871,
People's Republic of China}
\affiliation{Leibniz Institute for Astrophysics Potsdam, An der Sternwarte 16,
14482 Potsdam, Germany}

\correspondingauthor{Ting Li}
\email{liting@nssc.ac.cn}
\begin{abstract}

Large solar energetic particle (SEP) events are thought to originate from the shocks driven by fast coronal mass ejections (CMEs) and thus generally accompanied by type II radio bursts. However, a significant proportion of type II radio bursts is not accompanied by SEP events. To study the relationship between SEPs and type II radio bursts and the associated physical mechanisms, we statistically analyze 43 SEP halo-CMEs and 131 non-SEP halo-CMEs observed from 2010 to 2024, and check the related properties of type II radio bursts and solar source region. We find nearly all SEP events and approximately two-thirds of non-SEP events are accompanied by type II radio bursts. Type II radio bursts associated with SEP events usually have longer duration and lower ending frequencies. The starting frequency exhibits a clear source region dependence, being highest for “single active region (AR)”, intermediate for “multiple ARs”, and lowest for “outside of
ARs”. Furthermore, the spectra of both protons and electrons exhibit a similar softening trend in the three types of source regions. Joint analysis of spectra and type II radio bursts reveals that the proton spectra index has a good anti-correlation with the starting frequency of the type II radio bursts. Our statistical results have important implications for the mechanisms behind SEP acceleration.

\end{abstract}


\section{Introduction} \label{sec:intro}
Solar energetic particles (SEPs) are an important indicator of particle 
acceleration in solar eruptions. This phenomenon typically involves two acceleration scenarios: impulsive events accelerated a include disruptions t flare reconnection sites \citep{1986Cane,1999Reames} and gradual events caused by coronal mass ejection (CME)–driven
shocks \citep{1993Gosling,2003Kahler&Reames,2013Reames}. SEP events are hazardous to electronics and humans in space. These disruptions include damage in communications, navigation, spacecraft electronics, power systems and even commercial aviation\citep[e.g.,][]{2005Iucci,2008Cucinotta,2015Tobiska,2025Papaioannou}. As humans pursue space exploration outside of low Earth orbit, the understanding, monitoring, and forecasting of SEP events becomes increasingly important.

Type II radio bursts are also an important manifestation of particle acceleration. It is widely concluded that type II radio bursts are generated by electrons accelerated at coronal/interplanetary (IP) shocks \citep{1958Ginzburg&Zhelezniakov,1983H&P}. 
Specifically, the accelerated electron beams excite Langmuir waves due to beam plasma instabilities and the subsequent mixing with other wave modes in the corona, thus producing type II radio emissions \citep{1958Ginzburg&Zhelezniakov,1985Dulk}. 
In spectra, they appear as lanes with negative frequency drift observed between the metric and kilometric frequency range. These lanes have a frequency ratio of close to 2, representing emission at the fundamental and harmonic of
the plasma frequencies \citep{1985Nelson&Melrose,1985McLean&Labrum}.

Although type II bursts are generated from radiation of nonthermal
electrons, while the SEPs are accelerated protons, electrons and ions observed by spacecraft located in the heliosphere, many studies
have established a connection between these two phenomena \citep[e.g.,][]{2003Gopalswamy,2004Cliver&Kahler&Reames,2005Gopalswamy,2015Winter&Ledbetter}. \citet{2008Gopalswamy} studied 473 fast and wide CMEs during
1996 to 2005, and concluded that the SEP association rate of IP type II bursts originating from the western hemisphere of the
Sun is 100\% when the CME speed is $>$1800 km s$^{-1}$. Moreover, they found fast and wide CMEs lacking type II radio bursts are also absence of large SEP events, implying the close relation between type II burst and SEP events.
\citet{2016papaioannou} studied 174 SEP events covering solar cycle 23 and
the ascending phase of solar cycle 24 (1997–2013) and found a 74\% correlation between SEP events and
solar radio type II bursts in the decametric-hectometric (DH) range.
Historical research has shown that the close link between the SEP and type II bursts.

Many studies have focused on the specific links of type II bursts with SEP. Considering the wavelength, when type II bursts have emission components in all the spectral domains: metric, DH and kilometric (km), their association rate with SEPs shows a significant increase with broader wavelength \citep{2004Cliver&Kahler&Reames,2005Gopalswamy,2008Gopalswamy,2017Miteva}. This indicates the shock asscoiated with SEP is strong enough to produce radio emission over a large range of distances from the Sun. 
\citet{2017Prakash} conducted detailed investigations on type II bursts of SEP-associated and non-SEP-associated events observed during the period of 1997–2012. It is shown that SEP-associated type II bursts have higher frequency drift rate than those of the non-SEP-associated events, which reflects a higher shock speed in SEP-associated eruptions. Recently, \citet{2020Iwai} compared 26 spectral structures of type II bursts with corresponding SEPs. They found that the bandwidth of the hectometric type II bursts and the peak flux of the SEPs have a positive correlation. This result suggests that more SEPs may be generated from wider or stronger CME shock with a longer duration.

Energy spectra is another important aspect of SEP research which usually reflects the particle acceleration process. 
In the scenario of shock acceleration, the evolution of the shock-normal direction during its outward propagation has a significant impact on the spectra
\citep{2006T&L}. Some studies suggest quasi-perpendicular shocks can proceed much more harder spectral index than acceleration at parallel shocks \citep{2005Giacalone,2005Zank}.
In addition to shock acceleration, some studies proposed that flare acceleration
can contribute to the higher-energy portion of the SEP events \citep{2001K&T,2015Dierckxsens,2022Kiselev,2026Li}. 
However, few studies have incorporated the observation of source regions into analysis. \citet{2015Gopalswamy} reported four large filament eruptions (FEs) and found their softer spectra. They examined the starting frequency of the type II radio bursts and suggested that shock formation at larger distances from the Sun seems to be
the primary reason for the softer spectrum. Moreover, \citet{2025Duan} statistically analyzed the impact of source regions on SEP generation rates. \citet{2025Duan} classified all halo-CME events from 2010–2024 into three types: ``single AR'', ``multiple ARs'' and ``outside of ARs'' according to the relative location of their flare ribbons and active regions. In these cases, eruptions in ``multiple ARs'' and ``outside of ARs'' exhibit larger scales than those in ``single AR''. It is concluded that the SEP production rate is only about 20\% for events from ``single AR'', but for events associated with ``multiple ARs'' and for ``outside of ARs'', the production rate is around 35\%. This result indicates that the spatial scale of the source region has a significant impact on the production of SEPs.


In this paper, to address why a significant fraction of type II radio bursts occur without SEP events, we perform a statistical analysis of the properties of type II radio bursts. To follow the work of \citet{2025Duan} and provide more observational results, we also combine the source regions and spectra into our research.
This work is organized as follows: In Section \ref{sec:Data}, the process of statistically analyzing type II radio burst data and particle data is demonstrated. In Section \ref{sec:Results}, we present statistical results. At last, a summary and discussion of this work is provided in Section \ref{sec:Summary and Discussion}.
 
\section{Data and Methods} \label{sec:Data}
\begin{figure*}
    \centering
    \includegraphics[width=1.0\textwidth]{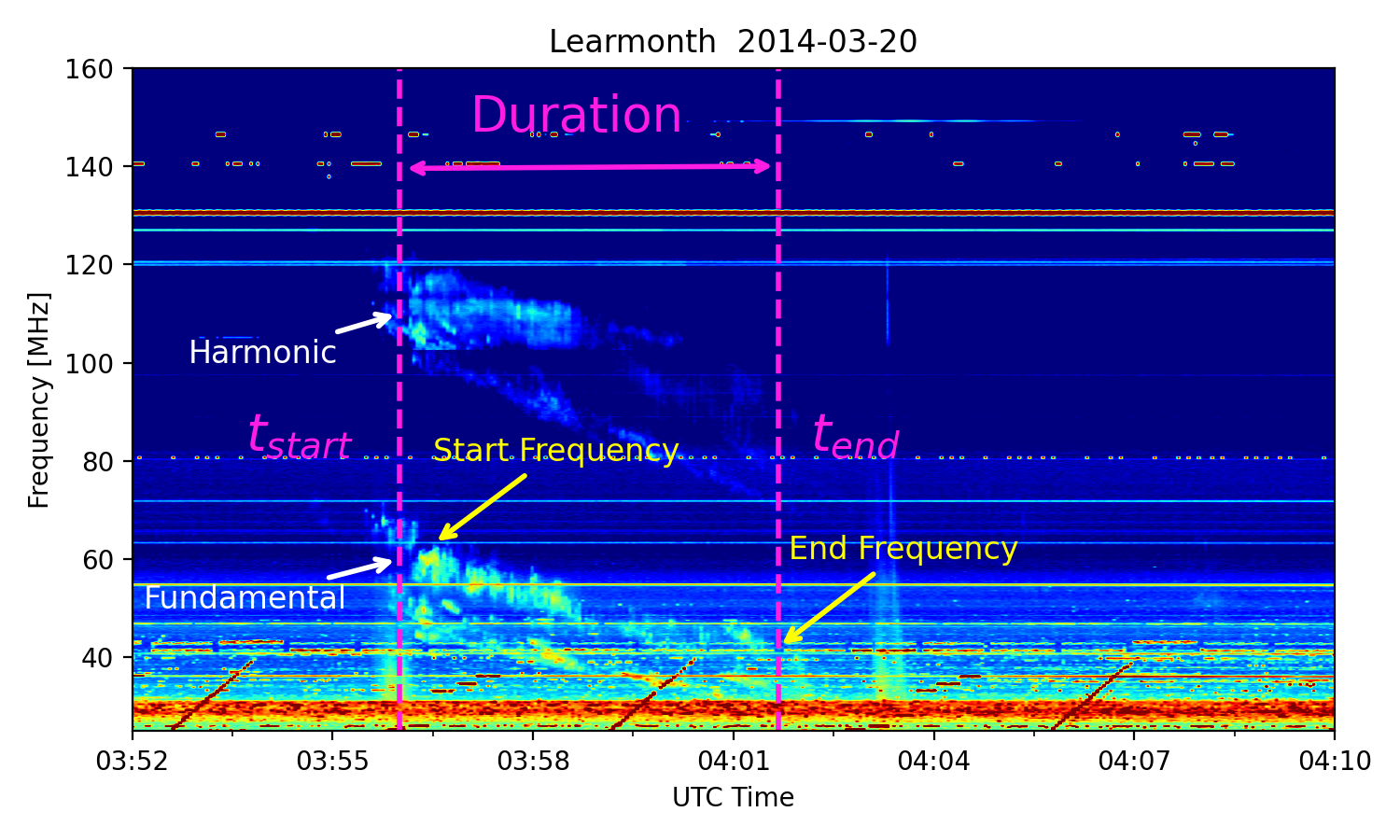}
    \caption{Solar radio dynamic spectra of a high-frequency type II burst on March 20, 2014. The start time, end time and duration of the burst are indicated by pink lines and arrows. The white arrows show the fundamental and harmonic bands in the spectra. Start and end frequency are shown in yellow arrows}
    \label{fig0}
\end{figure*}
We analyze 176 halo-CME events from 2010-2024 from the database of ``HCSEP'' of \citet{2025Duan}.
This database includes 45 SEP events and 131 non-SEP events from 2010-2024, which contains a variety of characteristics. Our catalog records the CME speed and flare class for all events. Moreover, this database classified solar sources of all halo-CME events into ``single AR'', ``multiple ARs'' and ``outside of ARs''. 
\subsection{Data of Type II Radio Burst}
For this study, we search the correspond type II bursts for 176 halo-CMEs from
``HCSEP''. It is widely recognized that type II radio bursts exhibit emission in different wavelength domains. A thorough investigation of their characteristics requires the synthesis of observations from multiple radio facilities. 
First, to identify the metric domains, we base our analysis on the type II radio bursts dynamic spectral data provided by the Radio Solar Telescope Network (RSTN)\footnote{\url{https://www.ngdc.noaa.gov/stp/space-weather/solar-data/solar-features/solar-radio/rstn-spectral/}}. The RSTN is a network of four identical solar observatories over the globe allowing for continuous observation
with frequency coverage from 25 to 180 MHz. A catalog of metric type II radio bursts based on RSTN is also utilized for our research \citep{2024Lawrance}\footnote{\url{https://catalogs.astro.bas.bg/.}}.  This catalog describes the m-type II associated solar flares
and CMEs as their origin in the solar corona.
To obtain radio data at frequencies above 180 MHz, we also utilize solar dynamic spectra taken with the ground-based Compound Astronomical Low cost Low frequency Instrument for Spectroscopy and Transportable Observatory (CALLISTO)\citep{2009Benz}. 
Second, after the metric domains is identified or there is no metric domains identified, we further complemented
their DH counterparts by using data from the Wind/WAVES instrument \citep{1995Bougeret}. We also utilize 
Wind/WAVES catalog\footnote{\url{ https://cdaw.gsfc.nasa.gov/CME_list/radio/waves_type2.html}}\citep{2019Gopalswamy}, maintained by the
Coordinated Data Analysis Workshop (CDAW) Data Center at
NASA and the Catholic University of America. The catalog compiles observations from the WAVES onboard the Wind spacecraft and the SWAVE instrument onboard the STEREO \citep{2008Bougeret}. 

Because the association time between solar flares and CMEs has been established and verified in ``HCSEP'', we only have to associate type II radio bursts with CMEs. To achieve this, the CME onset time and the start time $t_{start}$ of the type II burst are compared (up to 1-hr difference between type II with CME occurrences), we also use existing metric type II catalog for cross-checking the source region. Second, we try to determine the starting frequency, ending frequency and duration of each type II burst. This identification is typically conducted in the metric band, if both fundamental and harmonic frequencies are present, we take the starting frequency of the fundamental component as the starting frequency of the type II radio burst. To confirm and identify metric component of each event, we examine observational data from multiple sources, including RSTN and e-Callisto. If no sign of metric component is detected there, we search DH-counterparts using observations from Wind/Waves and SWAVES, and information based on their catalog (which contains explicit type II and CME correspondences). After the determination of the starting frequency, we also identify the ending frequency in both the metric and DH bands, and define the corresponding end time as $t_{end}$. If an event spans both the m and DH bands, we typically take the ending frequency of its last appearance in the DH band as the ending frequency for the entire event, and corresponding time as the end time. Finally, we calculate the duration ($t_{end}-t_{start}$) of the entire type II radio burst. 
We present a example of compiling the radio properties in Figure \ref{fig0}. In Figure \ref{fig0}, the starting frequency of this type II radio burst is around 60 MHz for fundamental component and the $t_{start}$ is around 03:56 UTC. The burst ends at a frequency of 40 MHz at around 04:01 UTC. We also examine the DH wavelength at this time and found no extension of this burst. Notably, for some events, the metric component is not always as easily distinguishable as shown in the Figure \ref{fig0}. For weak signals of type II bursts, we typically combine multiple data sources for confirmation. However, sometimes type II radio signals in the metric band can sometimes be submerged by type III bursts or other radio bursts. We also excluded events considering their signals in the metric band can sometimes be submerged by type III or other radio bursts, which usually makes it impossible to identify the actual type II start time and frequency. In our work, 2 halo-CME events are excluded from 176 halo-CME events in ``HCSEP''.

\subsection{Data of Proton and Electron}
In our work, we select 34 events with available proton and electron data for analysis. The remaining events are excluded from the analysis due to data gaps present in particle observations.
The energetic proton data is observed by the
Energetic and Relativistic Nuclei and Electron \citep[ERNE;][]{1995Torsti} instrument on board Solar Heliospheric Observatory (SOHO). In order to
get the energy spectra, we used the following ERNE energy
channels: (1) 13.0–16.0 MeV, (2) 16.0–20.0 MeV, (3) 20.0–25.0 MeV, (4) 25.0–32.0 MeV, (5) 32.0–40.0  MeV, (6) 40.0–50.0 MeV, (7) 50.0–64.0 MeV, (8) 64.0–80.0 MeV, (9) 80.0–100.0 MeV, (10) 100.0–130.0  MeV. It is important to note that in calculating the energy spectra, we only consider energy channels where the flux exceeded the background level. We have also excluded
the periods where the proton intensities peaked due to the Energetic Storm Particle event. Therefore, we filter for the first peak after the onset of the energetic particle event for every channel. The energetic electron data is is observed from the Electron, Proton, and Alpha Monitor \citep[EPAM;][]{1998Gold} instrument on the Advanced Composition
Explorer (ACE), with four energy channels: (1) 0.038-0.053 MeV, (2) 0.053-0.103 MeV, (3) 0.103-0.175 MeV, and (4) 0.175-0.315 MeV. The fitted spectra for proton and electron have the form $dN/dE=kE^{-\alpha}\ $, where N and E are the energetic particle density and energy, respectively, k is a constant, and $\alpha$ is the spectral index.
\section{Results} \label{sec:Results}
\subsection{The Overall Characteristics of Associated Type II Radio Bursts for SEP and non-SEP Halo CMEs}\label{}
\begin{figure*}
    \centering
    \includegraphics[width=1.0\textwidth]{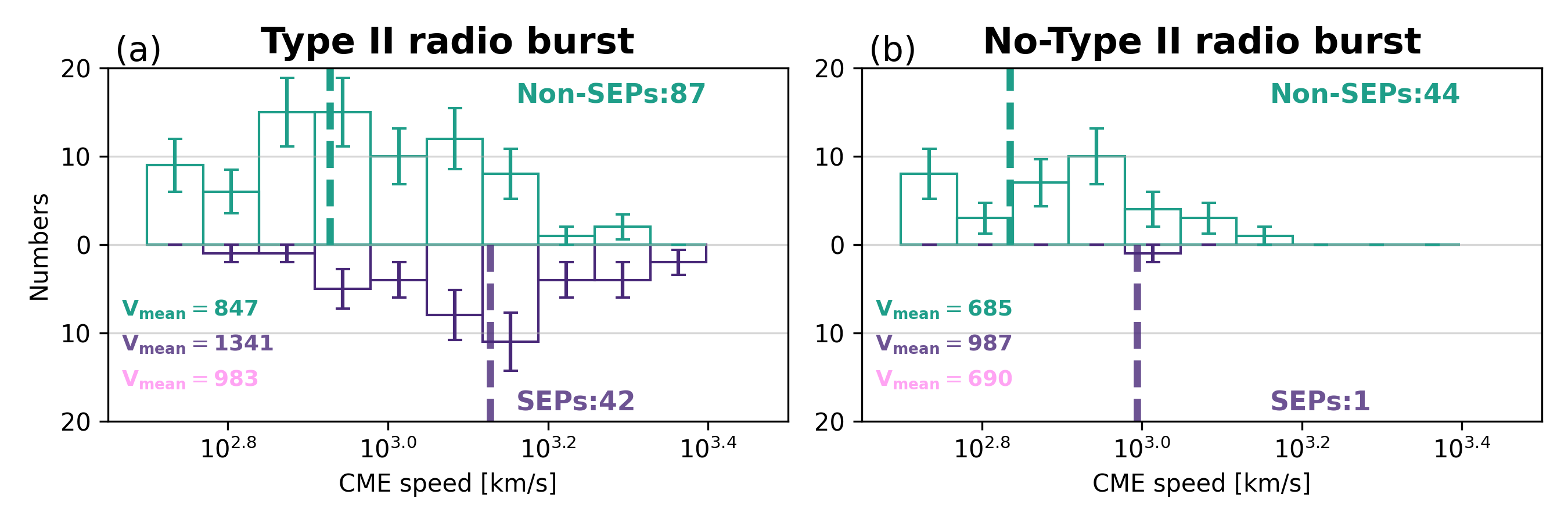}
    \caption{Distribution of halo-CME speeds for SEP (purple) and non-SEP (green) events. Panel (a): Type II radio bursts with 42 SEPs and 87 non-SEPs events; Panel (b): No-Type II radio bursts with 1 SEP and 44 non-SEPs events. Pink font denotes the overall mean CME speed for Type II and No-Type II subset. Mean speeds of SEP and non-SEP in each subset are indicated in dashed lines and fonts in purple and green. Each bar is accompanied by error bars that indicate the standard deviation.}
    \label{fig1}
\end{figure*}
Out of the 174 halo-CMEs in ``HCSEP'' with identifiable radio data, 129 type II radio bursts are identified in the radio dynamic spectra. This includes 42 SEP events and 87 non-SEP events. Among the 45 events that are not accompanied by type II bursts, there is only 1 SEP event and 44 non-SEP events.
For comparison, Figure \ref{fig1} presents halo-CME events in the ``HCSEP'' with and without type II radio bursts. In general, the halo-CME events associated with type II bursts have a higher average CME speed (983 km s$^{-1}$) than those without (690 km s$^{-1}$). Separately, among the 129 halo-CMEs accompanied by type II radio bursts (Figure \ref{fig1} (a)), SEP events exhibit significantly faster mean CME speeds (1341 km s$^{-1}$) compared with non-SEP events (847 km s$^{-1}$). Considering no type II subset (Figure\ref{fig1} (b)), although there is only 1 SEP event in no type II burst group, this SEP event still show higher speed (987 km s$^{-1}$) than the mean speed of rest non-SEP events (685 km s$^{-1}$). These results indicate two notable phenomena.
First, CMEs accompanied by type II radio bursts are significantly faster, indicating a close link between CME speed and shock formation. Second, 42 out of 43 SEP events are accompanied by type II radio bursts, which suggests that SEP production is closely related to shock acceleration. 
\begin{figure*}
    \centering
    \includegraphics[width=1.0\textwidth]{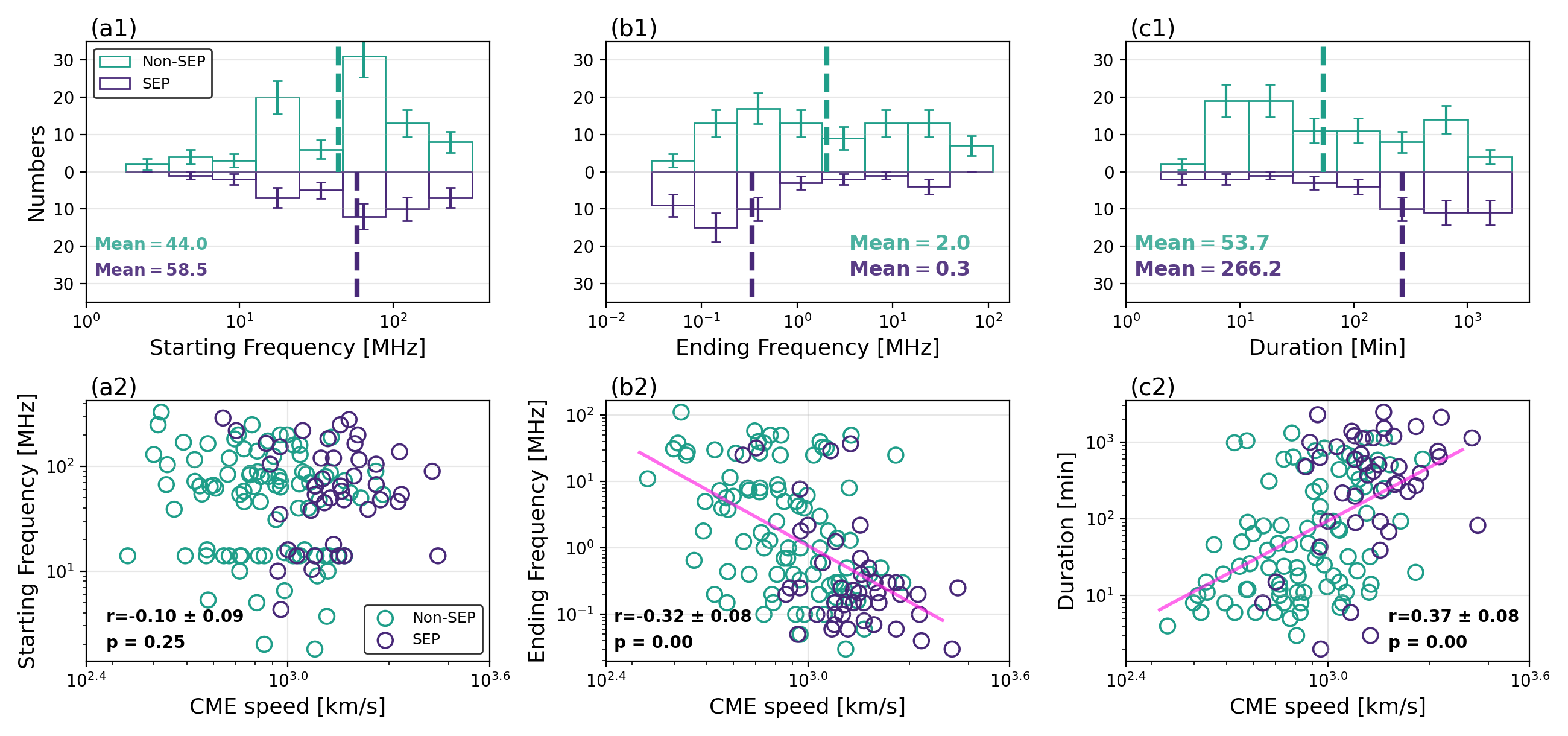}
    \caption{Statistical results for type II bursts properties (starting frequency, ending frequency, duration) and CME speeds. Panels (a1), (b1, (c1): Distribution of starting frequency, ending frequency, and duration of type II bursts for SEP and non-SEP events. Dashed lines and fonts in purple and green represent mean values for SEP and non-SEP events in different properties. Panels (a2), (b2, (c2): Scatter plots between CME speeds and the type II properties. The pearson correlation coefficients r and p values are provided. The errors in the correlation coefficients are also provided. Each bar is accompanied by error bars that indicate the standard deviation.}
    \label{fig2}
\end{figure*}

However, there are still 87 type II–associated halo-CMEs lacking detectable SEPs. To further investigate the differences between SEP and non-SEP events in type II subset of 129 halo-CMEs (Figure \ref{fig1} (a)), we conduct a statistical research of the type II radio burst properties in Figure \ref{fig2}. From the results of the starting frequency, we observe that both SEP and non-SEP events exhibit similar profiles (Figure \ref{fig2} (a1)), the peaks of both distributions are concentrated near 50 MHz. But the mean starting frequency of SEP events (59 MHz) is higher than that of non-SEP events (44 MHz). This implies the shock formation heights for SEP is closer to sun than non-SEP events. The ending frequency of type II bursts shows clear distinction between the SEP and non-SEP events (Figure \ref{fig2} (b1)). The ending frequency of SEP events is clearly concentrated at low frequencies, whereas non-SEP events show a more uniform distribution across the entire range. On average, SEP events have a lower mean value (0.3 MHz) than non-SEP events (2.0 MHz). It suggests that SEP-related shocks usually reach a larger distance from the Sun. Figure \ref{fig2} (c1) illustrates difference in the duration of type II bursts for SEP versus non-SEP events. SEP events display a notably longer type II burst duration than non-SEP events. The mean values also reveal a difference in duration between SEP (266.2 min) and non-SEP events (53.7 min). 
As seen from the scatter plot (Figure \ref{fig2} (a2)), there is a significant overlap between the SEP events and non-SEP events in starting frequency. Moreover, Considering all events, the starting frequency shows a weak correlation with CME speed. This indicates that CME speed has limited effect on shock formation height. In Figure \ref{fig2} (b2), it is clearly seen that SEP events correspond to the subset with higher CME speeds and lower ending frequencies (lower right part of the graph). Ending frequency of all events also exhibits a moderate negative correlation (r=-0.32) with CME speed.  Figure \ref{fig2} (c2) shows that SEP events correspond to the subset of longer duration and higher CME speeds (upper right part of the graph), with a moderate positive correlation (r=0.37) between the duration of all events and CME speed. 
From these results, we can conclude SEP-related type II bursts exhibit both lower ending frequencies and longer duration, implying SEP-associated shocks are shown to propagate farther and persist longer.
\begin{figure*}
    \centering
    \includegraphics[width=1.0\textwidth]{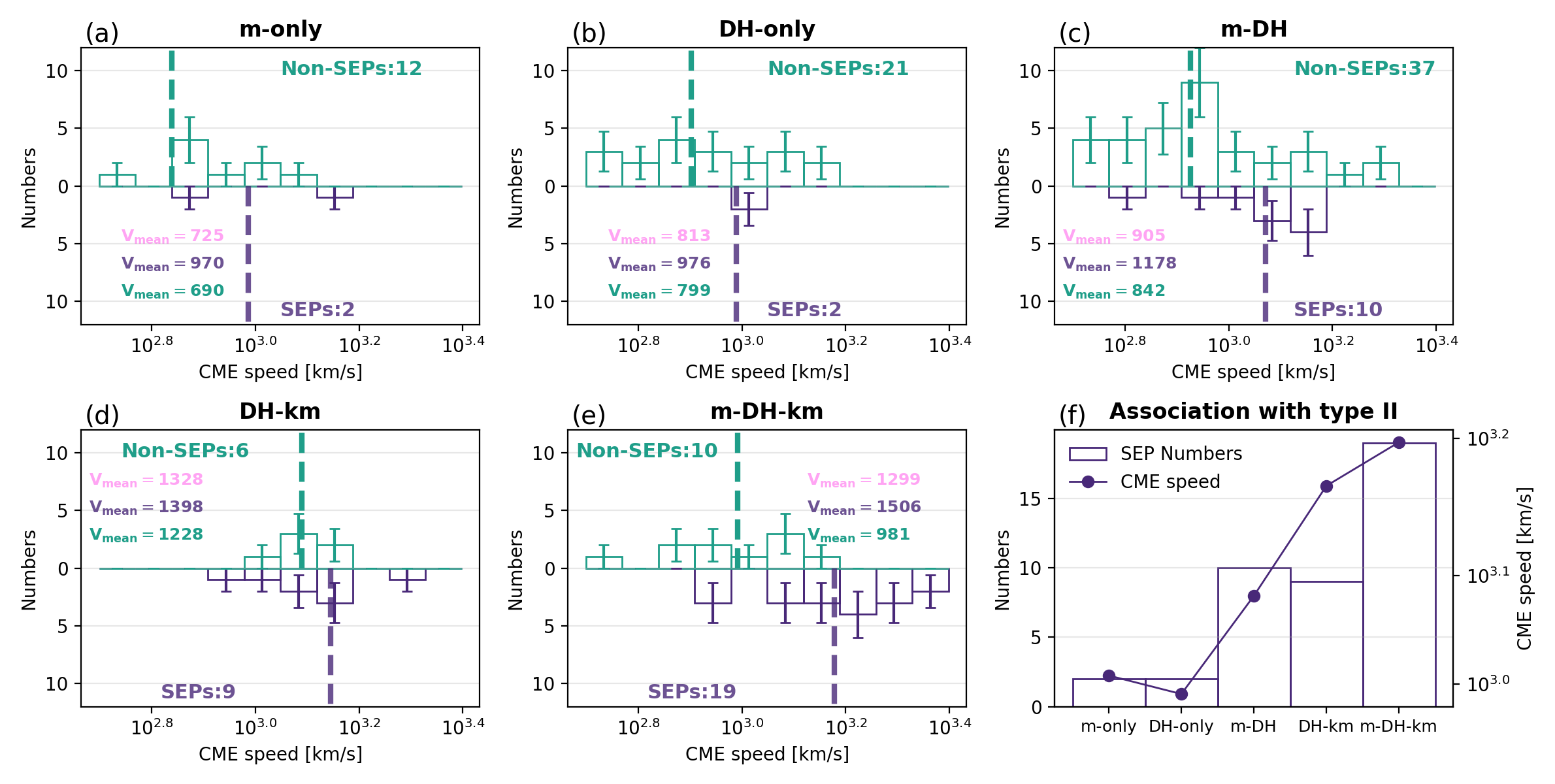}
    \caption{Statistical results for SEPs (purple) and non-SEPs (green) in different type II bursts categories. Panels (a)-(e): The number of SEPs and Non-SEPs in m-only, DH-only, m-DH, DH-km and m-DH-km subsets as function of CME speed. Dashed lines and fonts in purple and green represent mean values for SEP and non-SEP events. Pink font denotes the overall mean CME speed in different type II subset. Panel (f): The association between SEP numbers, CME speed, and the occurrence of type II bursts across these categories. Each bar is accompanied by error bars that indicate the standard deviation.}
    \label{fig3}
\end{figure*}

We further statistically categorize all 129 type II–associated halo-CMEs according to spectral domains based on their starting and ending frequencies. In Figure \ref{fig3}, we have shown the SEP events numbers and CME speeds for both SEP and non-SEP events in different sub-types of type II bursts. It is shown that type II bursts of m-only and DH-only (Figure \ref{fig3} (a) and (b)) sub-types exhibit a low SEP numbers (2 SEP events). But as the type II radio bursts have more spectral domains, the numbers of SEPs increases rapidly. SEP numbers are much higher in m-DH (10 SEP events) and DH-km (9 SEP events) subsets. And the SEP numbers peaks at the m-DH-km subset (19 SEP events).
Furthermore, average CME speed also increases from m-only to m-DH-km type II burst (Figures \ref{fig3} (a)-(e)). The average CME speeds are similar between the m-only and DH-only (725 km s$^{-1}$ and 813 km s$^{-1}$). As the spectral domains become broader, the average CME speed shows a slightly increase (905 km s$^{-1}$) in m-DH subset and rises significantly in the DH-km (1328 km s$^{-1}$) and m-DH-km (1299 km s$^{-1}$) subsets. Moreover, within each subset, the mean CME speeds increase for both SEP and non-SEP events. The average CME speeds are 970, 976, 1178, 1398 and 1506 km s$^{-1}$ for the SEP events in m-only, DH-only, m-DH, DH-km, m-DH-km subsets. For non-SEP events, the CME speeds are 690, 799, 842, 1228, 981 km s$^{-1}$, respectively. From Figure \ref{fig3} (f), we can see that the SEP numbers and CME speed rise as the type II radio bursts become broader in wavelength.

\subsection{Source Locations of SEP and non‑SEP Halo CMEs with Type II radio Bursts}\label{}
\begin{figure*}
    \centering
    \includegraphics[width=1.0\textwidth]{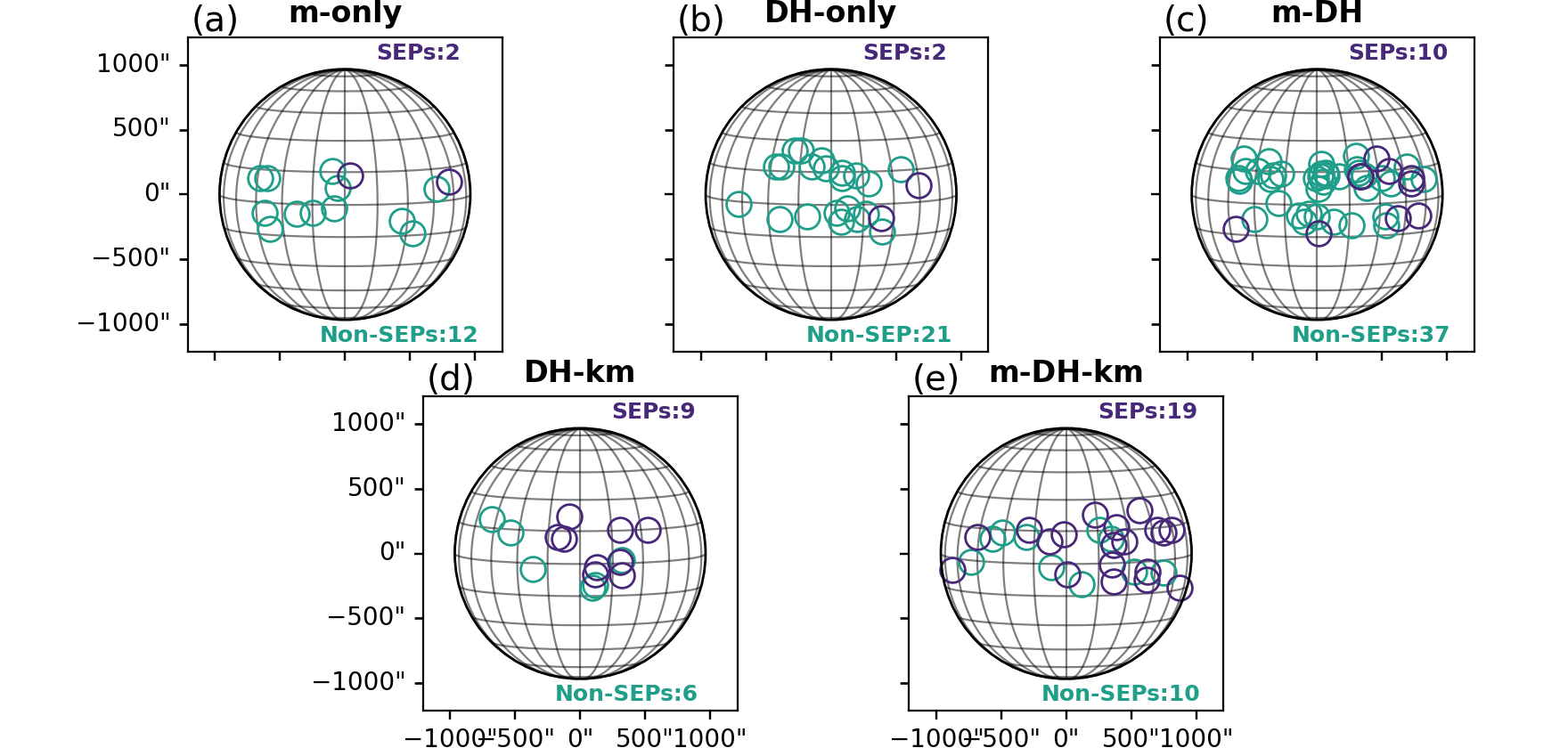}
    \caption{Location of SEP and non-SEP events in different type II radio bursts. Panels (a)-(e): m-only, DH-only, m-DH, DH-km, m-DH-km.}
    \label{fig4}
\end{figure*}

In Figure \ref{fig4}, we analyze the solar source regions for the 129 type II bursts. The location is evaluated by longitude and latitude on solar disk.
Figures \ref{fig4} (a) and (b) show that although there are only 2 SEP events in each of the m-only and DH-only subsets, all 4 SEP events are located in the western hemisphere. In m-DH subset, the number of SEP events increases (Figure \ref{fig4} (c)). 8 out of 10 events are located in the western hemisphere and only 2 events are located in the eastern hemisphere. But within the subsets of DH-km and m-DH-km (Figures \ref{fig4} (d)-(e)), we find that the occurrence of SEP events is not exclusively confined to the western hemisphere. 3 out of 9 SEP events in the DH-km category are located at central meridian. 6 out of 19 SEP events in the m-DH-km category are located at eastern hemisphere or central meridian. This indicates that the SEP source regions which generate farther propagating shocks are less sensitive to the magnetic connectivity. 
\subsection{Source Region Types of SEP and non-SEP Halo CMEs with Type II Radio Bursts and Spectral Index}\label{}
In Figure \ref{fig5}, based on the classification of \citet{2025Duan}, we investigate the type II radio properties within different source region types. We classify 129 type II–associated halo-CMEs into three groups: ``single AR'', ``multiple ARs'' and ``outside of ARs'' according to their size, structure, and location relative to ARs. From a comparison of the starting frequencies (Figures \ref{fig5} (a1)-(c1)), it is interesting to note that the mean values show a clear decreasing trend in the ``single AR'', ``multiple ARs'' and ``outside of ARs''. For all halo-CMEs regardless of SEP and non-SEP events, the highest mean value is 42.7 MHz for ``single AR'', followed by 38.4 MHz for ``multiple ARs'' and 32.2 MHz for ``outside of ARs''. Therefore, our results reveal that the eruptions in ``multiple ARs'', and ``outside of ARs'' generate shocks at a greater distance from the solar surface than those of ``single AR''. These results are consistent with the physical processes of solar eruptions. On one hand, events originate from ``multiple ARs'', and ``outside of ARs'' source regions have larger spatial scale. Thus, the CMEs they generate initiate from a higher coronal altitude, leading to the formation of their driven shocks at a greater height. On the other hand, the local Alfvén speed is inversely proportional to density. In higher coronal eruptions where the local density is lower, a CME requires a longer duration of acceleration to exceed the Alfvén speed, which consequently leads to the formation of a shock at a greater height.
For the three types of SEP-related events, the mean starting frequency is 58.5 MHz, 47.5 MHz and 38.5 MHz, which are higher than 38.6 MHz, 31.9 MHz and 26.8 MHz of non-SEP events. This indicates that the shock formation height associated with SEP events remains closer to sun in the three types than those of non-SEP events. 
In contrast, the ending frequencies of the three types are statistically similar in mean values (Figures \ref{fig5} (a2)-(c2)). The mean ending frequencies are 1.5 MHz, 0.5 MHz and 0.8 MHz for ``single AR'', ``multiple ARs'' and ``outside of ARs'', respectively. For SEP events in three types, the mean ending frequency is 0.4 MHz, 0.1 MHz and 0.4 MHz, which are lower than 2.1 MHz, 1.6 MHz and 1.5 MHz for non-SEP events. Across the three types of source regions, the ending frequencies of SEP events are still concentrated in the lower range of the statistical distribution, whereas that of non-SEP events are more broadly distributed.
In terms of type II duration, the mean values increase from ``single AR" (70.7 min) to ``outside of ARs" (127.2 min) and are highest for ``multiple ARs" (196.6 min). Across all source regions, SEP events still have a remarkable longer type II duration compared to non-SEP events. Within events of ``multiple ARs'', the mean value of type II duration in SEP events is 598.6 min, which are much higher than 74.2 of non-SEP events. Type II duration of SEP events in ``single AR'' and ``outside of ARs'' show mean values of 191.9 min and 268.8 min. In contrast, the mean values in non-SEP events are significant shorter in ``single AR'' (51.2 min) and ``outside of ARs'' (60.2 min).
\begin{figure*}
    \centering
    \includegraphics[width=1.0\textwidth]{MyFig/2Fig5.png}
    \caption{Statistical results of type II radio properties for SEP (purple) and non-SEP (green) events in three source region types. The count rate reflects the percentage of SEP and non-SEP events within different intervals. Panel (a1)-(a3): Start frequency, end frequency and duration of type II bursts for 21 SEP and 65 non‑SEP events in ``single AR''. Panel (b1)-(b3): The same for 7 SEP and 8 non‑SEP events in ``multiple ARs''. Panel (c1)-(c3): The same for 14 SEP and 14 non‑SEP events in ``outside of ARs''. Dashed lines and fonts in purple and green represent mean values for SEP and non-SEP events. Pink font denotes the overall mean values in different source region. Each bar is accompanied by error bars that indicate the standard deviation.}
    \label{fig5}
\end{figure*}

\begin{figure*}
    \centering
    \includegraphics[width=1.0\textwidth]{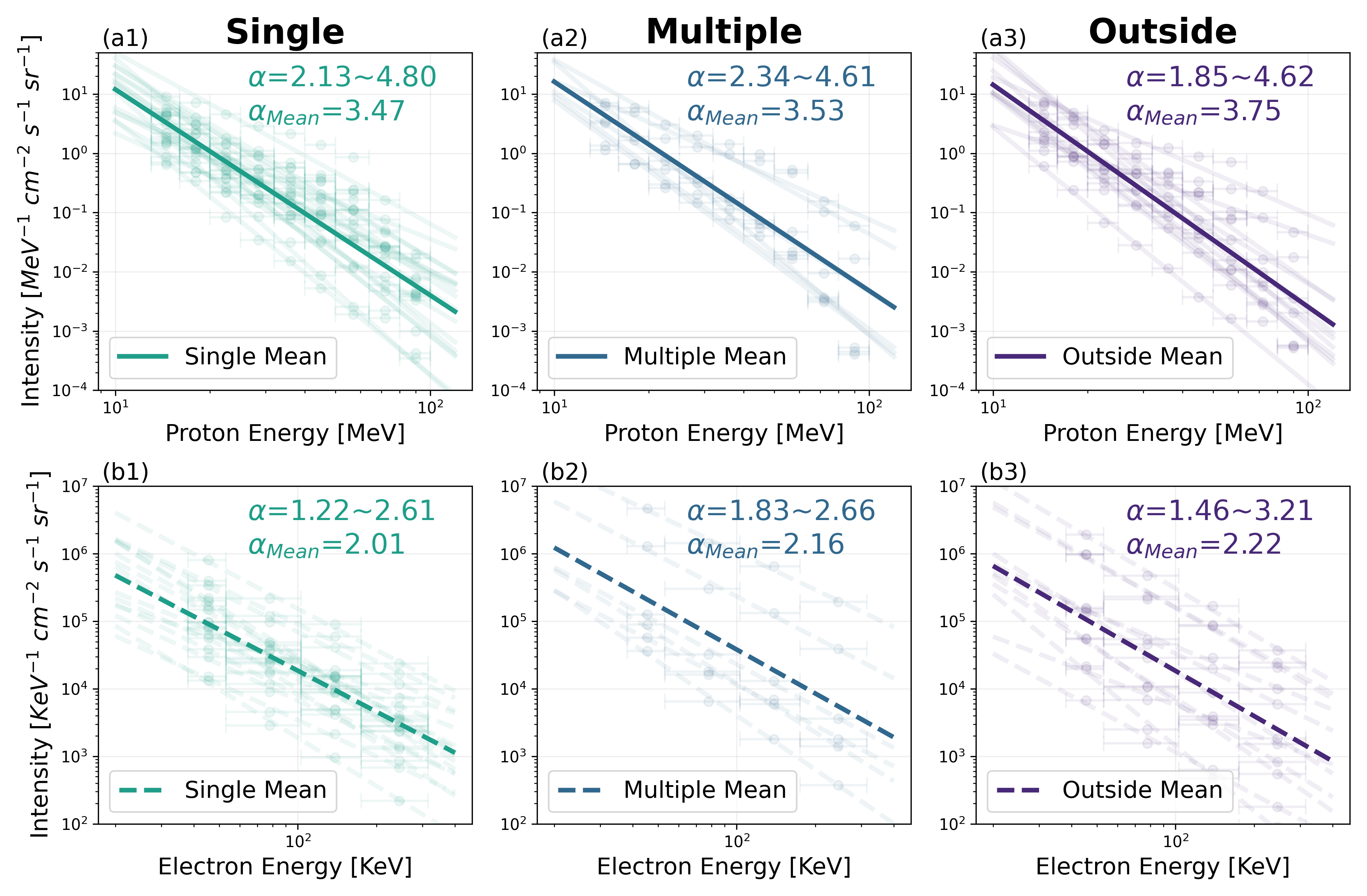}
    \caption{Comparison of proton and electron spectral index for SEP events originating from three source region types. Panels (a1)-(a3): Proton spectra in ``single AR'' (green), ``multiple ARs'' (blue) and ``outside of ARs'' (purple). Transparent scatter points with error bars represent the peak fluxes collected in different proton energy channels. The transparent lines denote the proton spectra fit for each event. The solid lines represent the average spectral index. Panels (b1)-(b3): The same as panels (a1)-(a3) but for electron spectra.}
    \label{fig6}
\end{figure*}

\begin{figure*}
    \centering
    \includegraphics[width=1.0\textwidth]{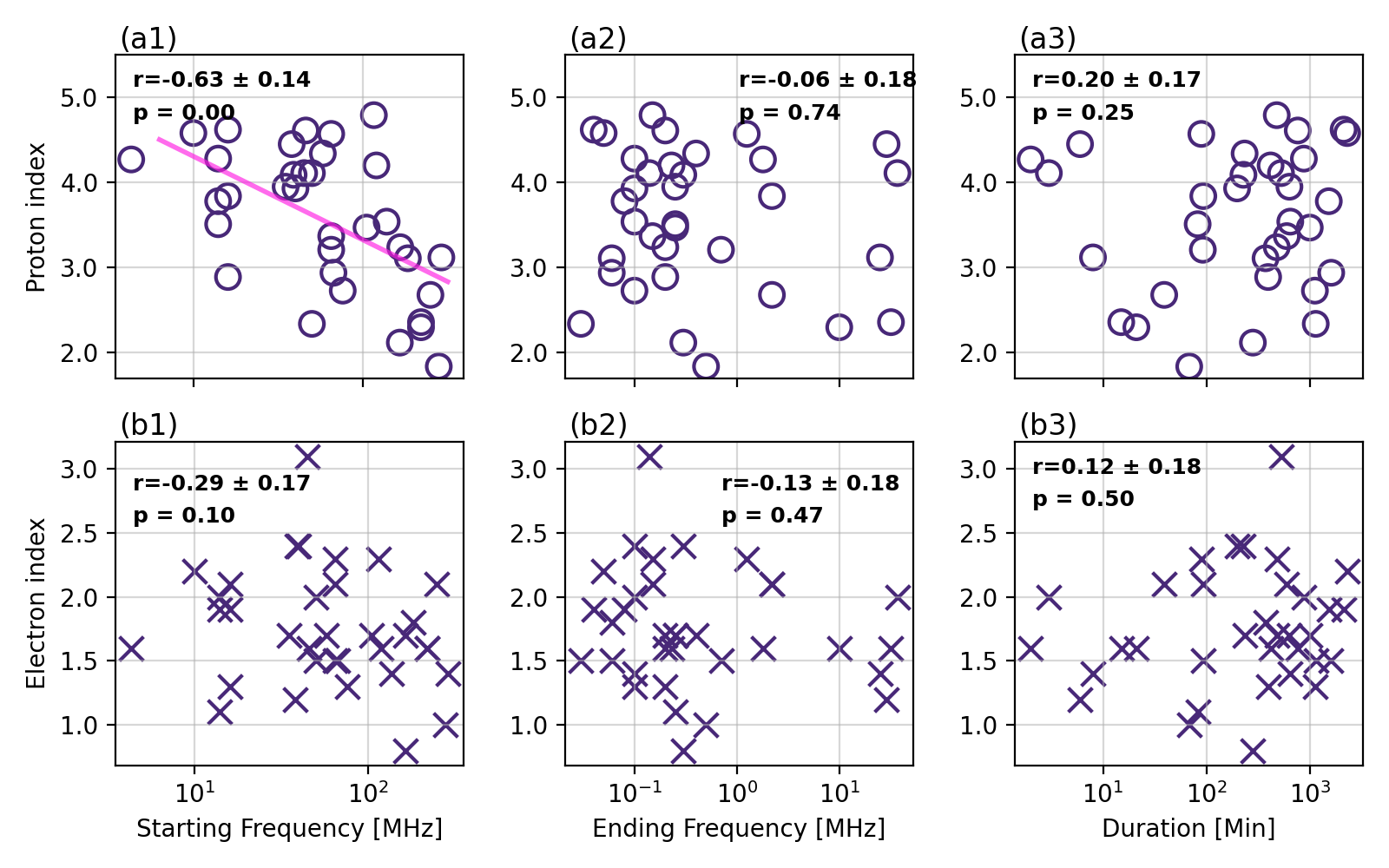}
    \caption{Statistical results of the proton spectra, electron spectra, and type II radio data for 34 SEP events. Panels (a1)-(a3): Scatter plots of the proton index (purple circle) as function of starting frequency, end frequency and duration. Panels (b1)-(b3): The same for electron index (purple cross). The pearson correlation coefficients r and p values are provided in each panel. The errors in the correlation coefficients are also provided.}
    \label{fig7}
\end{figure*}

The energy spectra of SEP are crucial for understanding the acceleration mechanism of energetic particles. To address this question, we analyze the proton and electron spectral index (power-law index) for our 34 SEP events with available proton and electron data. 
To investigate the impact of the source region on particle acceleration. we combine spectral index with source region classification.
Figure \ref{fig6} shows the background-subtracted time-of-maximum differential energy spectra of the 34 SEP events in different source regions for protons and
electrons. From the statistical analysis of the proton power-law index (Figures \ref{fig6} (a1)-(a3)), we can see that $\alpha$ in different source regions is different. The ``outside of ARs'' exhibits the widest range of $\alpha$ values (1.85 to 4.62). 
In terms of the mean value of $\alpha_{mean}$, we find that the spectra from ``single AR'' are the hardest ($\alpha_{mean}=3.47$), followed by 3.53 of  ``multiple ARs'' and softest for ``outside of ARs'' (3.75). This indicates that SEP events from ``multiple ARs'' and  ``outside of ARs'' produce less particles at high energies compared to ``single AR''. 
Turning to the statistical results for electrons (Figures \ref{fig7} (b1)-(b3)), the ``outside of ARs''  still exhibits the widest range of $\alpha$ values (1.46 to 3.21), followed by the ``multiple ARs'' (1.83 to 2.66) and finally the ``single AR'' (1.22 to 2.61). Moreover, we still find a trend for the $\alpha_{mean}$ to become softer from ``single AR'' (2.01) to ``multiple ARs'' (2.16) to ``outside of ARs'' (2.22).

In Figure \ref{fig7}, we show the scatter plots of proton and electron index. As shown in Figure \ref{fig7} (a1), we clearly find a strong negative correlation (r=-0.63) between the proton spectral index and the starting frequency. This implies a higher shock formation altitude leads to a softer energy spectra, meaning a smaller proportion of high-energy particles. A possible physical scenario is that for shock acceleration, its efficiency is closely related to the ambient magnetic field environment \citep{1994Kirk}. At higher shock starting heights, the magnetic field is weaker, resulting in fewer high-energy particles. Another possible reason is that earlier formed shock at a lower latitude has a longer acceleration distance, potentially enhancing efficiency. On the other hand, magnetic reconnection can provide channels for accelerated energetic particles to escape into interplanetary space \citep[e.g.,][]{2013Masson,2014Klein}. And these 
particles may subsequently be captured by shocks for further acceleration \citep[e.g.,][]{2012Mewaldt,2021Reames}. With regard to the shocks that form closer to the solar surface, they are more likely to capture such accelerated particles by magnetic reconnection and accelerate them to higher energies.

In contrast to the strong correlation found for protons, the relationship between the electron spectral index and the starting frequency is weak (r=-0.29) (Figure \ref{fig7} (b1)). We can infer that shocks may have limited effect on electron acceleration. 
When considering the ending frequency of the type II bursts (Figures \ref{fig7} (a2),(b2)), we find that both the proton and the electron spectral index show a poor correlation with it (r=-0.06 and r=-0.13). Similar poor correlations are also found in the duration of the type II burst with both the proton and electron spectral index (Figures \ref{fig7} (a3),(b3)).

\section{Summary and Discussion} \label{sec:Summary and Discussion}
In this paper, we attempt to identify differences in type II radio burst characteristics between SEP and non-SEP events. This analysis is combined with an examination of the source region location and classification. Therefore, we statistically analyze all 176 halo-CMEs events with 45 SEP events from 2010 to 2024 May from database ``HCSEP'' and 174 halo-CMEs events are used in our work. Specifically, type II bursts are recorded in 42 SEP events (2 SEP events are excluded). A further analysis demonstrates that the type II bursts associated with SEP events exhibit much lower ending frequencies and longer duration compared to non-SEP events. We classify these type II radio bursts based on their wavelength. It is shown that type II bursts with emission components in broader spectral domains have higher CME speed and are more likely to produce SEP events, specifically from m-DH-km subset. Considering the source location, SEP events associated with broader wavelength type II bursts are no longer limited to the western hemisphere. Through the analysis of source region classification, we find that the starting frequency of type II radio bursts for ``outside of ARs" (mainly from the large-scale intermittent filament eruption) is lower than that for ``single ARs". The proton and electron energy spectra associated with ``outside of ARs'' are softer than those from ``single AR''. A statistical analysis of 34 SEP events shows a good negative correlation (-0.63) between the proton spectral index and the starting frequency of type II radio bursts (i.e., a higher starting frequency corresponds to a harder proton spectra). But no significant correlation is observed for electrons index with starting frequencies.

Our analysis indicates that nearly all SEP events are accompanied by type II radio bursts. In our sample, about 90\% of SEP events show an association with type II bursts. This correlation is higher than 75\% and 74\% in previous studies \citep{2013Miteva,2016papaioannou}. The probable reason is that our sample focused on large SEP events, which are associated with fast and wide CMEs and therefore more likely to drive shocks \citep[e.g.,][]{2008Gopalswamy,2016Desai}. We further find the associated type II radio bursts of SEP events have lower ending frequencies and longer duration compared to non-SEP events. However, the starting frequencies show no significant difference between the two. Similar statistical results have been reported in \citet{2017Prakash}, who found that SEP and non-SEP events exhibit statistically difference in the end frequency and duration for their DH type II component based on 31 SEP and 17 non-SEP events. However, our work is based on a larger sample (42 SEP-associated type II bursts and 87 non-SEP-associated type II bursts). In our work, we include m‑only, DH‑only, and other type II bursts categories in the statistical analysis and we still find
significant difference between SEP-associated type II bursts and non-SEP-associated type II bursts in both ending frequency and duration. This indicates SEP-associated CME events are more energetic than
those not associated with SEPs within the larger sample.
In terms of the wavelength, the SEP-associated type II radio bursts cover boarder spectral domains, specially in m-DH-km subset. This finding is consistent with previous studies \citep[e.g.,][]{2005Gopalswamy,2017Miteva}. It is thought that ending frequencies reflects the kinetic energy of CME \citep{2005Gopalswamy,2006Gopalswamy}. 
In addition, \citet{2021Patel} present the statistic research of DH type II bursts for the Solar Cycles 23 and 24. They found duration of type II radio bursts also partly related with CME initial speed. 
In this scenario, we can infer that both the duration and ending frequency of type II radio bursts are closely related to the kinetic energy of the associated CME. It implies that the higher-kinetic-energy CMEs can drive strong shocks that are capable of persisting a large range of distances from the Sun, and these long-duration shocks are more effective particle accelerators. On the other hand, higher-kinetic-energy CMEs can generate shocks with higher compression ratios, which can accelerate particles more efficiently.
Considering the source location, SEP events associated with broader wavelength type II bursts are no longer limited to the western hemisphere. One possible reason is that the shocks surviving beyond high altitude are more likely to have broad longitudinal extents, enabling less well connected shocks to intercept open field lines connecting to Earth \citep{2004Cliver&Kahler&Reames}. In this scenario, the shocks can accelerate particles at a far distance from their source and results in more observed SEP events.
Analysis of the source regions reveals a systematic trend: the starting frequency of type II radio bursts is highest for events originating in ``single AR'', intermediate for ``multiple ARs'', and lowest for ``outside of ARs''. This reflects the difference in the formation heights of the shock from different source regions. 
According to the study of \citet{2025Duan}, eruptions from ``multiple ARs'' typically involve larger-scale structures that span multiple AR sizes. And eruptions from ``outside of ARs'' are often characterized by the eruption of large-scale intermittent filaments. Both of these source regions clearly involve a significantly larger scale eruption at high altitude. Thus, these eruptions naturally generate shocks at a higher altitude.

We statistically find that the proton spectral index shows a good anti-correlation with the starting frequency of type II radio bursts. This indicates a close link between proton acceleration and the shock formation height. For shock acceleration, the decay of the ambient magnetic field with distance, the longer acceleration distance at lower heights, and the secondary acceleration of high-energy particles captured by shocks at lower heights may collectively contribute to this phenomenon.
Previous studies revealed low starting frequency seems to be a common property of all the high-spectral-index events by investigating SEP events associated with filament eruptions \citep{2015Gopalswamy,2026Gopalswamy}. Our work not only confirms this trend but also identifies that eruptions originating from ``multiple ARs'' also correspond to similarly low starting frequencies and softer spectra index. This suggests that the spatial scale of the source region to some extent affects the shock formation height. 
Specifically, among these three types, the average shock formation height is highest for ``outside of ARs'', intermediate for ``multiple ARs'', and lowest for events from ``single AR''. Correspondingly, their average spectral index of proton $\alpha_{mean}$ become progressively softer in the same order. Our results show a close connection linking the spatial scale of the source regions to the characteristics of the energy spectra.
On the other hand, considering the contribution of flares to proton acceleration may also explain the phenomenon of the proton energy spectra varying with the source region. In a previous study, \citet{2001K&T} proposed that low energy protons are mainly accelerated by shocks driven by CMEs, while high-energy protons primarily originate from flare processes. This interpretation was further supported by several independent studies \citep{2015Dierckxsens,2017RAA,2026Li}. For eruptions of ``outside of ARs'', the corresponding flare classes are often weaker and this may also lead to fewer high-energy protons.

Considering the electron energy spectra, we find little correlation with the starting frequency. This may indicates a difference in the acceleration mechanisms between electrons and protons. 
In our study, the electron energy range we use($<$0.4 MeV) belongs to the near-relativistic electrons. Relevant studies have shown that multiple acceleration processes may contribute to acceleration of near-relativistic electrons  \citep[e.g.,][]{2015Trottet,2015Koulo,2021Dresing}. Moreover, recent studies have shown that there is no clear correlation between the spectral index of near-relativistic electrons and shock parameters \citep{2022Dresing,2023Rodr}. Therefore, the limited role of the shock in near-relativistic electron acceleration may explain the weak correlation between electrons and shock formation heights in our results. 

\section*{Acknowledgments}
We thank the open data of RSTN, e-Callisto network,
SOHO, ACE, STEREO and Wind instruments.
This work is supported by the B-type Strategic Priority Program of the Chinese Academy of Sciences (grant No. XDB0560000), the National Natural Science Foundations of China (12533010,12333009, 12273060), the National Key R\&D Programs of China (2022YFF0503800), and the Youth Innovation Promotion Association of CAS (2023063). This work is supported by China's Space Origins Exploration Program (GJ11020405), and the Specialized Research Fund for State Key Laboratory of Solar Activity and Space Weather.
\bibliography{main}
\bibliographystyle{aasjournal}
\end{document}